# Accurate theoretical bandgap calculations of II-VI semiconductors

Imad Khan[1], Iftikhar Ahmad[1,*], H. A. Rahnamaye Aliabad[2], M. Maqbool[3]

1. Department of Physics, University of Malakand Chakdara, Pakistan
2. Department of Physics, Sabzevar Tarbiat Moallem University, Sabzevar, Iran
3. Department of Physics and Astronomy, Ball State University, Muncine, IN, USA

Corresponding Author: Professor and Chair, Department of Physics
University of Malakand, Chakdara, Dir(L), Pakistan
ahma5532@gmail.com 0092-332-906-7866

In this letter we present band gaps of II-VI semiconductors, calculated by the full potential linearized augmented plane wave (FP-LAPW) method with the modified Becke-Johnson (mBJ) potential. The accuracy of the calculated results is assessed by comparing them with the experimentally measured values. After careful analysis of the results presented in this paper, we found that the mBJ potential is very efficient in the predication of the bandgaps of II-VI semiconductors. It is also revealed that the effectiveness of mBJ is based on the proper treatment of the d-orbitals in the highly correlated electron system.

Density functional theory (DFT) has proven its worth in the past as an effective/ leading theoretical technique for the calculation of various physical properties of solids, while in present it is unmatchable in accuracy and applicability and in the future it is expected to grow further in all dimensions. The Kohn-Sham equations [1)] are extensively solved with the local density approximation (LDA) [1,2)] and generalized gradient approximation (GGA) [3)] for the structural, electronic, optical, magnetic and other physical properties of metals, semimetals, semiconductors, insulators, superconductors etc. Though, these calculations are effective for certain substances, but are ineffective in the calculations of the band structures of the highly correlated electron systems, with d or f orbital like II-VI compounds.

The II–VI semiconductors have been extensively studied due to their effective use in optoelectronic industry. These compounds are commonly used in many established commercial electronic and optoelectronic devices operating in blue to ultraviolet spectral regions such as visual displays, high-density optical memories, transparent conductors, solid-state laser devices, photodetectors, solar cells etc. These compounds crystallize in zinc-blende (B3), wurtzite (B4) and rock salt (B1) structures [4–7]. The knowledge of the optical properties of these materials is essential for the design and analysis of II-VI based optoelectronic devices. Therefore, the optical properties of these compounds have widely been studied experimentally as well as theoretically, and extensive information on the subject is available in literature.

As II–VI semiconductors are widely used in many optical and optoelectronic devices, hence their understanding is extremely important. Most of the theoretical results of these compounds are too far from the experimentally measured values, and hence they are unreliable. The band structures and the electronic properties of these compounds are investigated by local density approximation (LDA), generalize gradient approximation (GGA) and GW[8,9]. The calculated values of the bandgaps of the II-VI compounds by these techniques are presented in Table 1. The comparison of the results obtained by these techniques with the experimentally measured values confirm that the errors in most of the calculated results are more than 10%, which is technically not acceptable in the scientific community. Though, GW is some what better than the other two techniques, but the problem with GW is expensive calculations.

In the case of II–VI materials, it is important to include the localized ‘d’ orbitals in the cation. The localized ‘d’ orbitals play important role in the bonding process; hence their inclusion as valence orbitals is essential for a correct band structure [46] and optical spectra. LDA and GGA not only underestimate bandgaps but also band dispersions; particularly the location of d energy level come out incorrectly [4]. Thus the reason of the ineffectiveness of these techniques, especially the most commonly used LDA and GGA, is their inefficient treatment of the d state electrons.

In the present work we study the electronic structure of MX (M = Zn, Cd and X = O, S, Se, Te) compounds in the rock-salt, zinc-blende and wurtzite structures with the modified Becke and Johnson (mBJ)[47] exchange potential in the framework of full-potential linearized augmented plane wave (FP-LAWP) method as implemented in the WIEN2K package [48]. The mBJ potential is given as follow and the detail can found in Refs [34,47,49].

$$\nu_{x,\sigma}^{MBJ}(r) = c\,\nu_{x,\sigma}^{BR}(r) + (3c-2)\frac{1}{\pi}\sqrt{\frac{5}{12}}\sqrt{\frac{2t_{\sigma}(r)}{\rho_{\sigma}(r)}}$$

Where $\rho_\sigma = \sum_{i=1}^{N_\sigma} |\psi_{i,\sigma}|^2$ is the electron density, $t_\sigma = (1/2)\sum_{i=1}^{N_\sigma} \nabla\psi^*_{i,\sigma}.\nabla\psi_{i,\sigma}$ is the kinetic energy density and $\nu^{BR}_{x,\sigma}$ is the Becke-Roussel (BR) potential [50)].

For the calculations, $R_{MT's}$ are chosen in such a way that there was no charge leakage from the core and hence total energy convergence was ensured. For wave function in the interstitial region the plane wave cut-off value of $K_{max}=7/R_{MT}$ was taken and fine k mesh of 2000 was used in the Brillouin zone integration and convergence was checked through self-consistency. The convergence was ensured for less than 1 mRy/a.u.

The calculated bandgaps of the II-VI semiconductors with mBJ are presented in Table 1. To assess the accuracy of the calculated bandgaps, they are compared with the experimental results and are also compared with the other theoretical calculations in the table. It is obvious from the table that the results obtained by mBJ are very close to the experimental results.

The experimental bandgap of ZnO in the zinc blende phase is 3.27eV, while in our calculations it is 3.15 eV. The table shows serious underestimation and overestimation for the same phase of ZnO by LDA, GGA and GW methods. Similarly, for wurtzite ZnO our calculated bandgap value is very close to the experimental one, whereas LDA severely underestimates and GW overestimates the result. In case of ZnS we have excellent results for B3 and B4 phase. For ZnSe we have accurate results in B3 phase but a little underestimation for B4 phase. The calculated results for CdX (O, S, Se and Te) by mBJ are in good agreement with the experimental values as compared to the other calculated results. Most of the calculated bandgaps for MX by mBJ are almost the same as the experimental ones, only a few of them give error of less than10 % from the experimental results.

To compare the accurateness of the different theoretical techniques and visualize the effectiveness of mBJ; experimental and calculated bandgaps of the zincblend and wurtzite ZnO, ZnS, ZnSe, ZnTe, CdS, CdSe and CdTe are plotted in Figs. 1. and 2, respectively. It is clear from the plots that LDA is a poor technique for the calculation of the bandgaps of chalcogenides. It severely underestimates the bandgaps and for some compounds this underestimation could be more than 50 %. The figures also show that though GW is comparatively better than LDA but even then, for the most of chalcogenides it overestimates, while in the case of the zinc blende ZnO it severely underestimates the bandgap; whereas, the plots clearly show that the results provided by mBJ are very close to the experimental ones. The accurate results of the binary chalcogenides by mBJ predict, that this technique can also be used for the bandgap engineering of II-VI semiconductors.

The origin of a band structure of a compound is related to the corresponding density of states [27,51]. Any potential which is effective in treating the electronic states will be also effective in the band structure of a compound. The total and partial densities of states (DOS) for ZnO in the zinc blende phase are compared for mBJ and LDA in Fig. 3. The figure reveals that the effectiveness of the mBJ potential is due to its proper treatment of the electronic states as compared to other techniques. The mBJ enhances the intensity of the Zn-3d state as compared to LDA and it also pushes Zn-3d state towards the Fermi level. This shift leads to better theoretical treatment of the band structures and optical properties (closer to experimental result) of a compounds as compare to other techniques.

In summery we calculated the band structures of theoretically challenging II-VI semiconductors by using the modified Becke and Johnson (mBJ) potential within the full-potential linearized augmented plane wave (FP-LAWP) method. The calculated results are compared with the experimental and other theoretical results. It is concluded that mBJ is an efficient theoretical technique for the calculation of the band structures of chalcogenides (II-VI). The technique is much superior to the commonly used LDA, GGA, GW and other theoretical methods. It is also revealed that the efficient results provided by mBJ are due to the proper treatment of the electronic states. The results predict that mBJ will be a successful tool for the bandgap engineering of II-VI compounds.

**Figure Captions**

**Fig. 1.** (color online). Theoretical versus experimental bandgaps of the zinc blende mono-chalcogenides

**Fig. 2.** (color online). Theoretical versus experimental bandgaps of the wurtzite mono-chalcogenides

**Fig. 3.** (color online). Comparison of the total and partial densities of states of mBJ with LDA

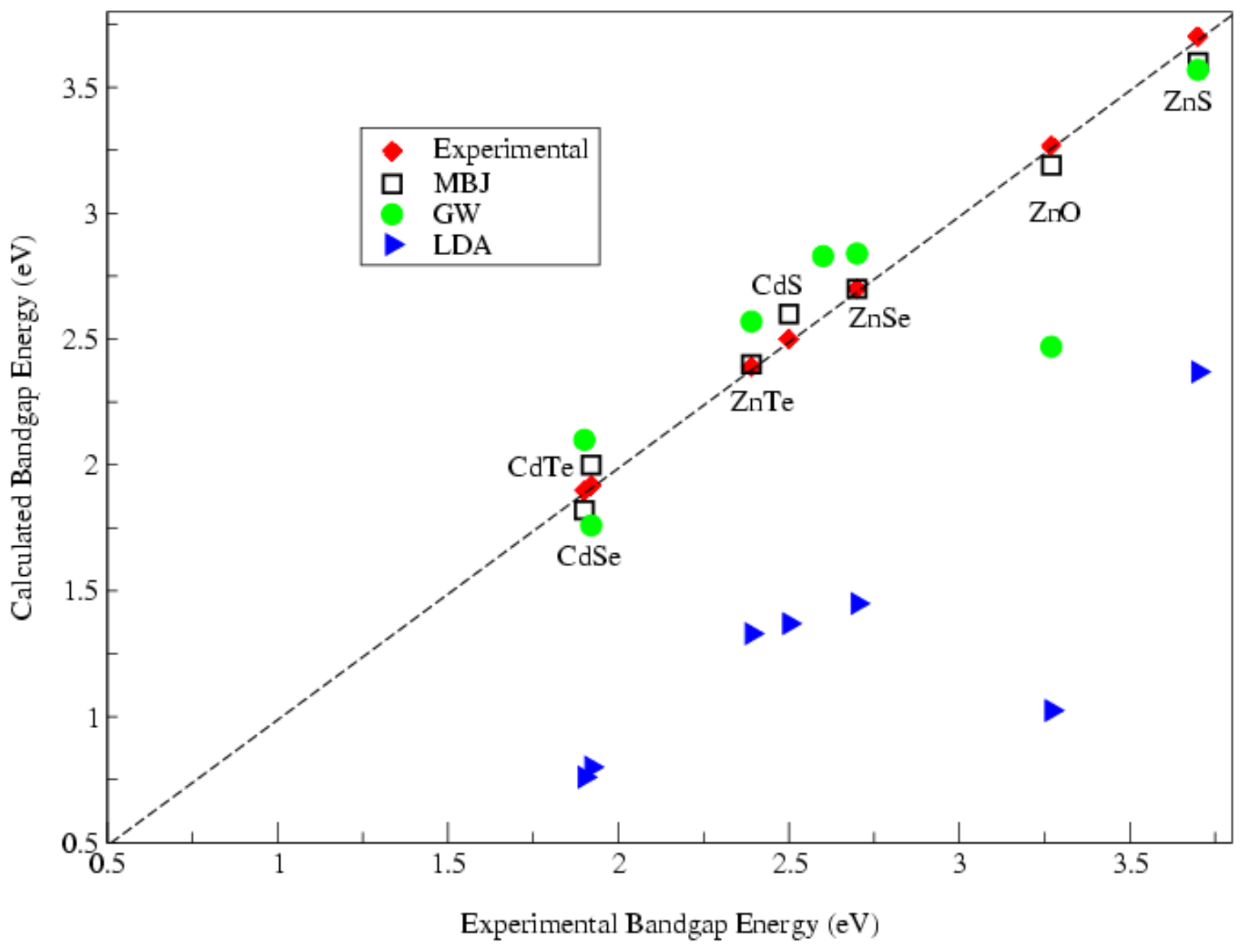

**Fig. 1.** (color online). Theoretical versus experimental bandgaps of the zinc blende mono-chalcogenides

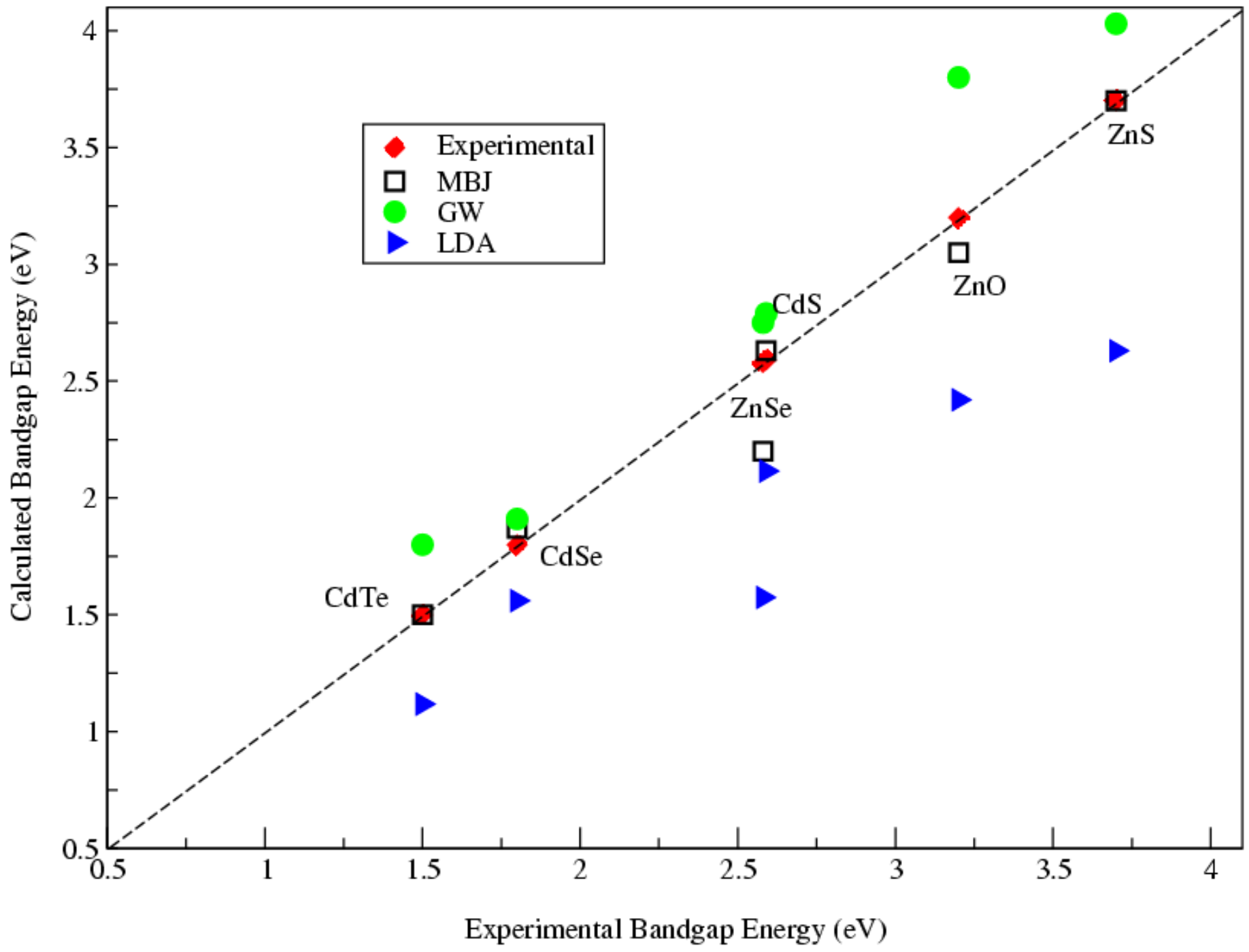

**Fig. 2.** (color online). Theoretical versus experimental bandgaps of the wurtzite mono-chalcogenides

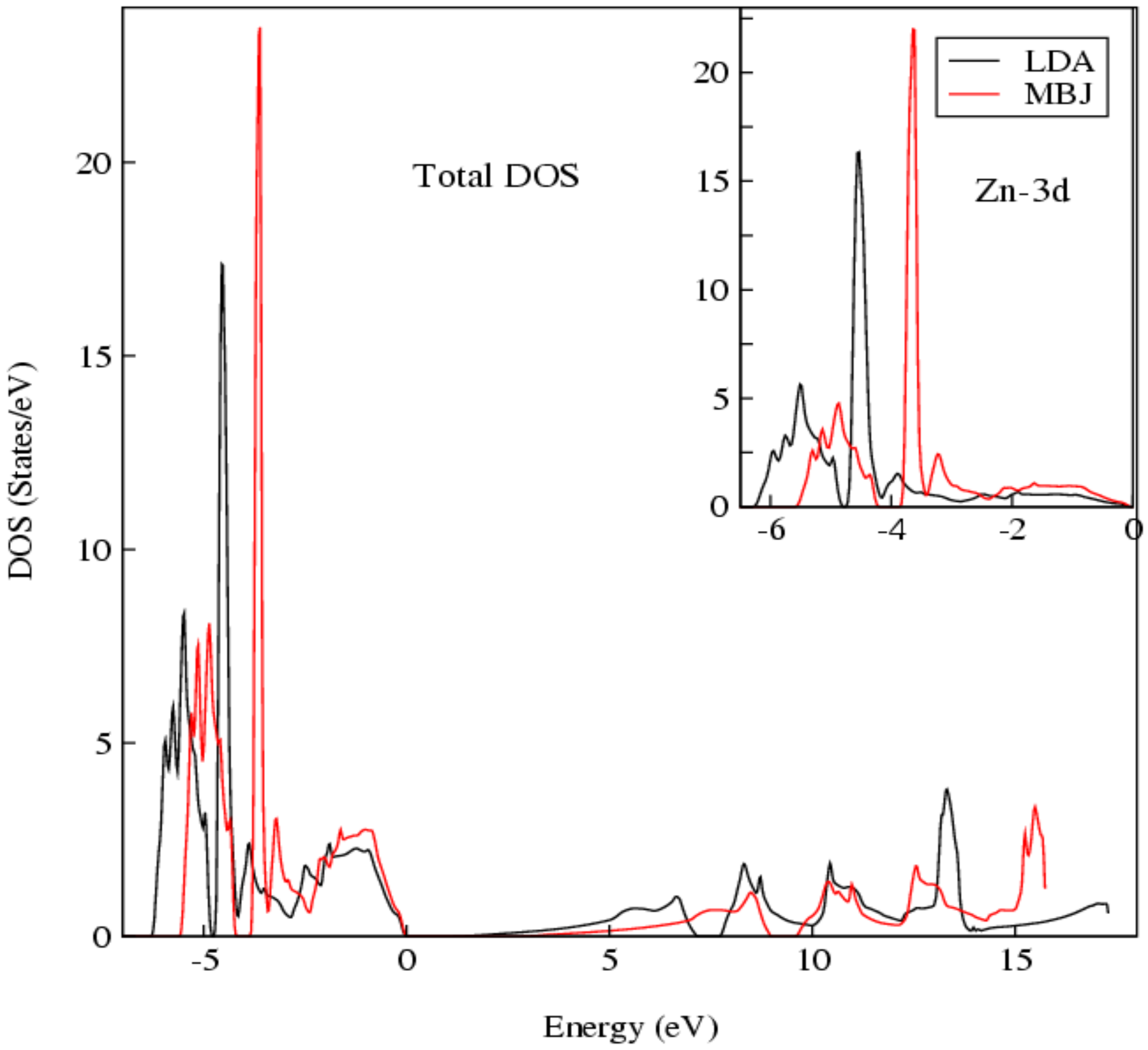

**Fig. 3.** (color online). Comparison of the total and partial densities of states of mBJ with LDA

**Table 1:** Theoretical and experimental fundamental bandbgaps (in eV) for II-VI compounds

| | Phase | Present | Experimental | GW | LDA/ GGA |
|---|---|---|---|---|---|
| ZnO | B 1 | 2.5 | 2.30±0.15[h] | 1.16[a], | 1.1[aj], 5.54[n] |
| | B 3 | 3.15/ 2.7[b] | 3.27[a] | 1.0[a],2.47[o] ,3.59[j] | 0.79[a], 3.5[b], 0.8[b], 0.71[q], 0.8[r], 1.025[x] |
| | B 4 | 3.05/ 2.68[ac] | 3.2[am], 3.3[p], 3.37[f], 3.4[e] | 3.8[ad], 2.44[an], 3.2[a] | 2.42[i], 0.81[q], 1.838[y], 2.367[y] |
| ZnS | B 1 | 1.63 | | | 0[ag], 2.9[p] |
| | B 3 | 3.6/ 3.7[b],3.66[ac] | 3.1[t], 3.7[b], 3.8[o] | 2.2[m] , ,3.97[j], 3.50[l], 3.98[k], 3.57[o], 4.15[ad] | 1.8[b], 3.6[b], 2.14[q], 2.11[r], 1.981[x], 2.37[k] , 2.016[g] |
| | B 4 | 3.7 | 3.86[j] | 4.03[k], | 2.24[y], 2.63[y], 2.45[k] |
| ZnSe | B 1 | 0.85 | < 1.1[ae, r] | | 1.1[aa], 1.16[25] |
| | B 3 | 2.7 | 2.7[s], 2.69[t] | 3.10[j], 2.84[k] | 1.45[k] |
| | B4 | 2.2 | 2.87[j], 2.58[z] | 2.75[k] | 1.092[y], 1.574[y], 1.43[k] |
| ZnTe | B 1 | -0.93 | | | |
| | B 3 | 2.4 | 2.39[t],2.3[w] | 2.57[k] | 1.163[x], 1.33[k] |
| | B 4 | 2.1 | | 2.67[k] | 1.092[y], 1.785[y], 1.48[k] |
| CdO | B 1 | 1.55 | 0.84[u], 1.09[ag], 1.98[ao] | | 0.83[ag], 2.70[ag], 1.04[m], 1.16[m], 0.85[v], 0.50[ag], 1.01[ag] |
| | B 3 | 2.63 | | | |
| | B 4 | 2.23 | | 1.06[aq] | |
| CdS | B 1 | 1.4 | | | |
| | B 3 | 2.56/ 2.7[b], 2.66[ac] | 2.5[b] | 2.83[k], 2.87[ad] | 0.9[b], 3.0[b], 1.37[k] |
| | B 4 | 2.63 | 2.49[t], 2.53[15], 2.59[ab] | 2.79[k] | 1.315[y], 2.115[y], 1.36[k] |
| CdSe | B 1 | 0.78 | | 0.7[af] | |
| | B 3 | 1.84 | 1.8[al],1.90[ab] | 2.01[k] | 0.76[k] |
| | B 4 | 1.87 | 1.8[al], 1.74[t], 1.74[z], 1.97[ab] | 1.91[k] | 0.78[y], 1.56[y], 0.75[k] |
| CdTe | B 1 | -0.56 | | | |
| | B 3 | 1.68 | 1.43[t], 1.5[w], 1.92[ab] | 1.76[k] | 0.80[k] |
| | B 4 | 1.5 | 1.5[z] | 1.80[k] | 0.644[y], 1.118[y], 0.85[k] |

[a]Ref. 10, [b]Ref. 11, [c]Ref., [e]Ref. 12, [f]Ref. 13, [g]Ref. 14, [h]Ref. 15, [i]Ref. 16, [j]Ref. 17, [k]Ref. 18, [l]Ref. 19, [m]Ref. 7, [n]Ref. 20, [o]Ref. 21, [p]Ref. 22, [q]Ref. 23, [r]Ref. 24, [s]Ref. 25, [t]Ref. 26 [u]Ref. 27, [v]Ref. 28, [w]Ref. 29, [x]Ref. 0, [y]Ref. 31, [z]Ref. 32, [aa]Ref. 33, [ab]Ref. 34, [ac]Ref. 35, [ad]Ref. 36 [ae]Ref. 37, [af]Ref. 38, [ag]Ref. 39, [aj]Ref. 40, [al]Ref. 41, [am]Ref. 42, [an]Ref. 43, [ao]Ref. 44, [aq]Ref. 45